\documentclass[final,3p,times,twocolumn]{elsarticle}

\usepackage{amssymb,amsmath,amsthm,bm}
\usepackage{geometry}
\usepackage{algorithm}
\usepackage{algpseudocode}

\usepackage{graphicx}
\graphicspath{{figures-pdf/}}
\usepackage[normalem]{ulem}
\usepackage{url}
\usepackage{lineno}

\usepackage{bm}
\usepackage{graphicx,color,overpic}

\usepackage{float}

\newlength\imagewidth
\newlength\figwidth
\newlength\sfigwidth
\newlength\vfigskip
\journal{IEEE Multimedia}
\usepackage{color}
\definecolor{dgreen}{rgb}{0,.6,0}

\usepackage[font=small,skip=0pt]{caption}
\emergencystretch=\maxdimen
\usepackage{hyperref}
\hypersetup{
 linktocpage=true,
 pdfborderstyle={/S/S/W 1},
 hyperindex=true,
 bookmarks=true,
 bookmarksopen=true,
 bookmarksnumbered=true,
}

\begin{document}

\begin{frontmatter}


\title{Cryptanalyzing an Image Scrambling Encryption Algorithm of Pixel Bits}

\author[xtu-cn]{Chengqing Li\corref{corr}}
\ead{DrChengqingLi@gmail.com}

\author[xtu-cn]{Dongdong Lin}

\author[cas-cn]{Jinhu L\"u}

\cortext[corr]{Corresponding author.}

\address[xtu-cn]{
Hunan Province Cooperative Innovation Center for Wind Power Equipment and Energy Conversion,\\
College of Information Engineering, Xiangtan University, Xiangtan 411105, Hunan, China}

\address[cas-cn]{Academy of Mathematics and Systems Sciences, Chinese Academy of Sciences, Beijing 100190, China}

\begin{abstract}
Position scrambling (permutation) is widely used in multimedia encryption schemes and some international encryption standards, such as DES and AES.
This paper re-evaluates the security of a typical image scrambling encryption algorithm (ISEA). Using the internal correlation remaining in the cipher-image, we can
disclose some important visual information of the corresponding plain-image under the scenario of a ciphertext-only attack. Furthermore, we have found the real \textit{scrambling domain},
position-scrambling scope of the scrambled elements of ISEA, which can be used to support an efficient known/chosen-plaintext attack on it. Detailed experimental
results have verified these points and demonstrated that some advanced multimedia processing techniques can facilitate the cryptanalysis of multimedia encryption algorithms.
\end{abstract}
\end{frontmatter}

\section{Introduction}

Position scrambling (permutation) is one of the simplest and most efficient methods for protecting all kinds of multimedia data \cite{Licq:hierarchical:SP2016,Cqli:Fridrich:SP2017}. More importantly, it can effectively facilitate an encryption scheme to obtain an efficient combination of the properties of confusion and diffusion. Over the past two decades, researchers have proposed many multimedia scrambling algorithms, designing different mechanisms to derive the position-scrambling relation of the scrambling elements from a secret key. Because the number of possible scrambling relations is the factorial of the number of scrambling elements, there are numerous different scrambling relations in theory when the number of scrambling elements is sufficiently large. As for image, video, and audio (speech), various forms of datum can be selected as scrambling elements, including the bit, pixel, and compressing coefficients of an image or video \cite{Li:Permutation:SPIC2008,WangQX:HDDCS:TCAS2016} or the transform coefficients and motion vectors of video \cite{Zeng:scrambling:TM2003} or audio frames \cite{Li:HDSP:IEE2006}. If the scrambling elements are spatial blocks of an image, the recovery of the original image can be classified as part of the general problem of solving jigsaw puzzles discussed in \cite{Liuhr:Jigsaw:TM2011}.

In contrast to cryptography, which is designing an algorithm to realize secure communication between the sender and the intended recipient, the object of cryptanalysis is to gain as much information about the secret key as possible without any prior knowledge of it \cite{Li:hyperchaotic:ND2013,Yap:ND:2015,Gex:Hide:IJBC16}. 

Some specific multimedia scrambling encryption algorithms have been successfully cryptanalyzed under different conditions, in terms of the number of required plaintexts, the computational complexity of the attack, and the storage space needed \cite{Arroyo:Permutation:SP13,Jolfaei:Permutation:IFS2016,Hua2016IS}. As
different scrambling elements may have dramatically different effects on the sensible information of the plaintext, different
multimedia scrambling encryption algorithms may possess totally different strengths against ciphertext-only attack.
However, their strengths against plaintext attacks can be evaluated by a uniform model \cite{Li:Permutation:SPIC2008}.
As cryptanalyzed in \cite{Li:Permutation:SPIC2008}, any multimedia scrambling encryption algorithm can be efficiently broken
with $O(\lceil \log_L(MN) \rceil)$ plaintexts and $O(\lceil\log_L(MN)\rceil\cdot MN^2)$ computational time, where
$MN$ is the size of the \textit{scrambling domain}, the position-scrambling scope of the scrambled elements, $L$ is the number of different value levels of the scrambled elements,
and $\lceil \cdot \rceil$ returns the smallest integer greater than or equal to a given number. In \cite{Lcq:Optimal:SP11}, the computational load
is further reduced to $O(\lceil\log_L(MN)\rceil\cdot MN)$ by replacing the operations of intersecting sets of position candidates with
linear visits of a branch tree whose node stores information about the position candidates.

In \cite{Ye:Scramble:PRL10}, a typical image scrambling encryption algorithm (ISEA) was proposed that scrambles a binary presentation
of a gray-scale plain-image with a pseudo-random number sequence generated by iterating a digital chaotic map. In 2011, ISEA was analyzed as
a scrambling encryption algorithm acting on a scrambling domain of size $M\times (8N)$ \cite{Lcq:Optimal:SP11}. In \cite[Sec.~3.1]{Ye:Scramble:PRL10}, a weaker version of ISEA was
suggested to decrease the computation complexity and save running time: every row and column of the binary matrix of size $M\times (8N)$ are scrambled
with the same scrambling vector. In 2012, a set of specific plaintexts were constructed by Zhao \textit{et al.} to recover the equivalent version of the secret key of
ISEA \cite{Zhao:scrambling:CNSNS2012}.

Here, we re-evaluate the security of the simpler version of ISEA and find the real reasons behind its attack advantage \cite{Zhao:scrambling:CNSNS2012}: it does not act on a scrambling domain of size $M\times (8N)$, but simply cascades two scrambling algorithms with scrambling domains of sizes $M$ and $8N$. We therefore propose an efficient known-plaintext attack and a general chosen-plaintext attack on the algorithm. We also found that important visual information of the plain image can be observed from a single cipher image encrypted by ISEA.

The rest of this paper is organized as follows. The next section briefly introduces
ISEA. The security of ISEA against ciphertext-only attack and plaintext attacks are re-evaluated in Sec.~\ref{cryptanalysis:ISEA} with some
experimental results. The last section concludes the paper.

\section{Description of ISEA}
\label{sec:encryptscheme}

The encryption object of ISEA is a gray-scale image of size $M\times N$
(height$\times$width), which can be represented as a matrix over the set $\{0, 1, \cdots, 255\}$:
$\bm{I}=[I(i, j)]_{i=0, j=0}^{M-1, N-1}$. The image $\bm{I}$ is further decomposed as an $M\times
(8N)$ binary matrix $\bm{B}=[B(i,l)]_{i=0, l=0}^{M-1, 8N-1}$, where
\begin{equation}
\sum_{k=0}^7 B(i, l)\cdot 2^k=I(i, j),
\end{equation}
$l=8\cdot j+k$. After preforming a scrambling of
the row and column vectors of $\bm{B}$, the cipher-image $\bm{I}'=[I'(i,j)]_{i=0, j=0}^{M-1,
N-1}$ is obtained, where $I'(i, j)=\sum_{k=0}^7 B'(i, 8\cdot j+k)\cdot 2^k$.
The basic concrete parts of ISEA can be described as follows.
\begin{itemize}
\item \textit{The secret key}: three positive integers $m$, $n$, and $T_i$, and
the initial condition $x_0\in (0,1)$ and control parameter $\mu
\in(3.569945672, 4)$ of the Logistic map
\begin{equation}
f(x)=\mu\cdot x\cdot(1-x).
\label{equation:Logistic}
\end{equation}

\item \textit{Initialization}: 1) run the Logistic map
from $x_0$ to generate a chaotic sequence, $\{x_k\}_{k=1}^{L}$, where $L=\max{\{(m+M),(n+8N)\}}$;
2) produce a vector $\bm{T_M}$ of length $M$, where $\bm{S_M}(\bm{T_M}(i))$ is the
$(i+1)$-th largest element of $\bm{S_M}=\{x_{m+k}\}_{k=1}^{M}$,
$i\in \{0, 1, \cdots, M-1\}$; 3) produce a matrix $\bm{T_N}$ of size
$1\times (8N)$, where $\bm{S_N}(\bm{T_N}(j))$ is the $(j+1)$-th largest element of
$\bm{S_N}=\{x_{n+k}\}_{k=1}^{8N}$, $j\in \{0, 1, \cdots, 8\cdot N-1\}$.

\item \textit{The encryption procedure}:

\begin{itemize}
\item \textit{Step 1 -- vertical permutation}: generate an intermediate matrix
$\bm{B}^*=[B^*(i,l)]_{i=0, l=0}^{M-1, 8N-1}$, where
\begin{equation}
B^*(i,:)=B(\bm{T_M}(i),:).
\label{eq:encrypt_ver}
\end{equation}

\item \textit{Step 2 -- horizontal permutation}: generate an intermediate matrix
$\bm{B}'=[B'(i,l)]_{i=0, l=0}^{M-1, 8N-1}$, where
\begin{equation}
B'(:, l)=B^*(:,\bm{T_N}(l)).
\label{eq:encrypt_hor}
\end{equation}

\item \textit{Step 3 -- repetition}: reset the value of $x_0$ with the current state of map~(\ref{equation:Logistic}), and repeat
the above operations from the \textit{initialization} part $(T_i-1)$ times.
\end{itemize}

\item \textit{The decryption procedure} is similar to the encryption except for some simple modifications:
1) the multiple rounds of encryption are executed in the reverse order;
2) \textit{Step 2} is executed first in each round;
3) the left parts and right parts of Eq.~(\ref{eq:encrypt_ver}) and Eq.~(\ref{eq:encrypt_hor})
are swapped, respectively.
\end{itemize}

Note that the full version of ISEA is given in \cite[Sec. 2]{Lcq:Optimal:SP11}, where
the matrix $\bm{T_N}$ is of size $M\times (8N)$, each row vector $B^*(i, :)$ is scrambled by
the $i$-th row vector of $\bm{T_N}$. As the full version of ISEA was cryptanalyzed comprehensively in
\cite{Lcq:Optimal:SP11}, only the simpler version is studied in the remainder of this paper.

\section{Cryptanalysis of ISEA}
\label{cryptanalysis:ISEA}

\subsection{Ciphertext-only Attack on ISEA}

The ciphertext-only attack is an attack model for cryptanalysis where the attacker can only access some ciphertexts.
Obviously, possessing sufficient robustness to withstand the attack is a basic requirement for any multimedia encryption algorithm.
Unfortunately, ISEA fails to satisfy the requirement.

The essential form of the encryption process of ISEA can
be represented by
\begin{equation*}
\bm{B}'=(\bm{T}_L)^{T_i}\cdot \bm{B} \cdot (\bm{T}_R)^{T_i},
\end{equation*}
where $\bm{T}_L$ and $\bm{T}_R$ are permutation matrices representing the vertical permutation in Eq.~(\ref{eq:encrypt_ver}) and the horizontal permutation in Eq.~(\ref{eq:encrypt_hor}), respectively \cite{Zhao:scrambling:CNSNS2012}.
As is well known, the multiplication product of any number of arbitrary permutation matrices is still a permutation matrix, a square binary matrix possessing exactly one entry of 1's in each row and each column. Regardless of
the value of $T_i$ and whether the intermediate matrices $\bm{B}^*$ in Eq.~(\ref{eq:encrypt_ver}) and $\bm{B}'$ in Eq.~(\ref{eq:encrypt_hor}) are updated in each round, the final essential form of the encryption process of ISEA is still
\begin{equation}
\bm{B}'=(\bm{\hat{T}}_L)^{T_i}\cdot \bm{B} \cdot (\bm{\hat{T}}_R)^{T_i},
\end{equation}
where $\bm{\hat{T}}_L$ and $\bm{\hat{T}}_R$ are permutation matrices. So, the repetition of the operations in Step 3) have no influence on obtaining
the equivalent version of the secret key of ISEA.
Without loss of generality, we just set the number of repetitions $T_i=1$ in the following cryptanalysis.

\begin{figure}[!htb]
\centering
\begin{minipage}[t]{\imagewidth}
\centering
\includegraphics[width=\imagewidth]{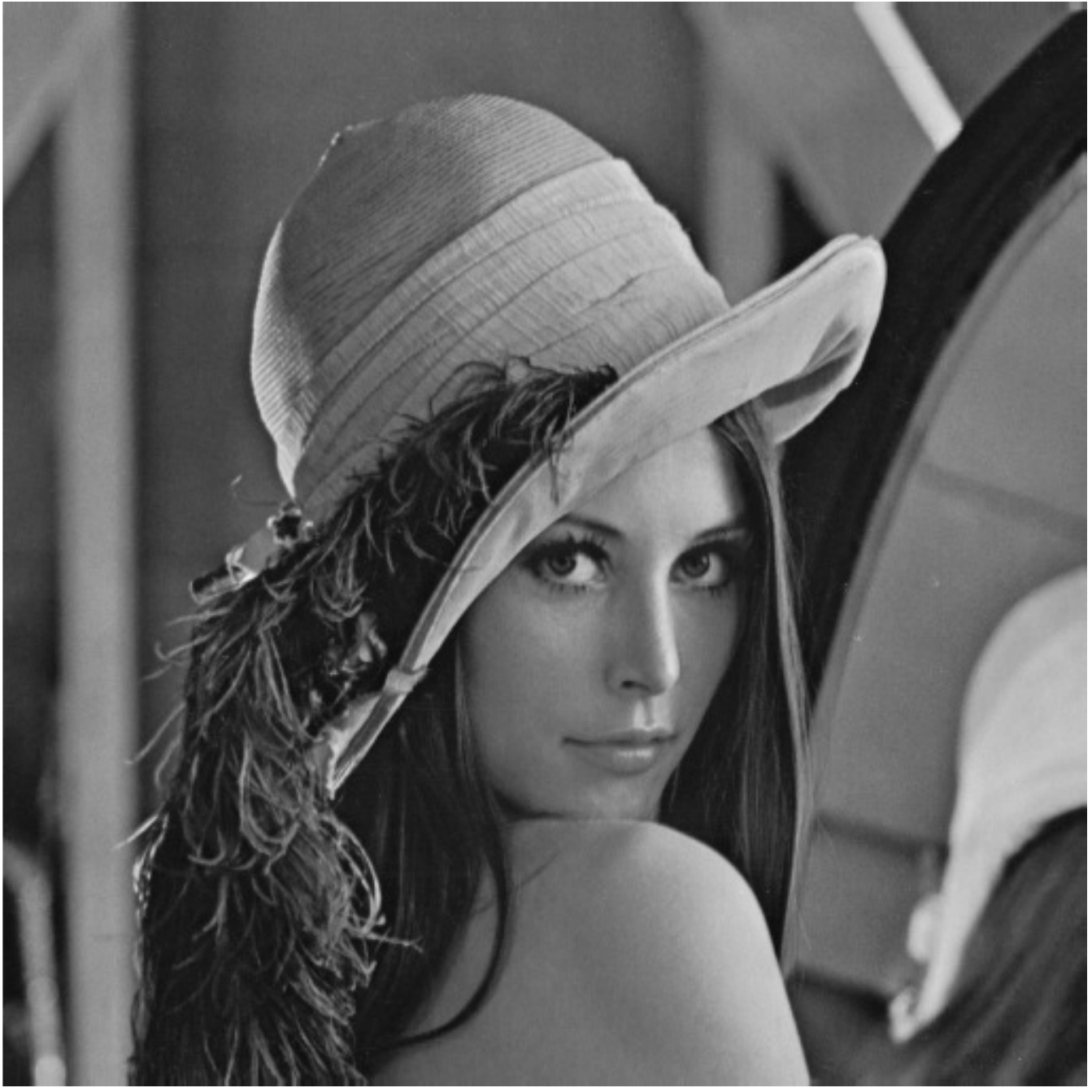}
a)
\end{minipage} \hspace{4pt}
\begin{minipage}[t]{\imagewidth}
\centering
\includegraphics[width=\imagewidth]{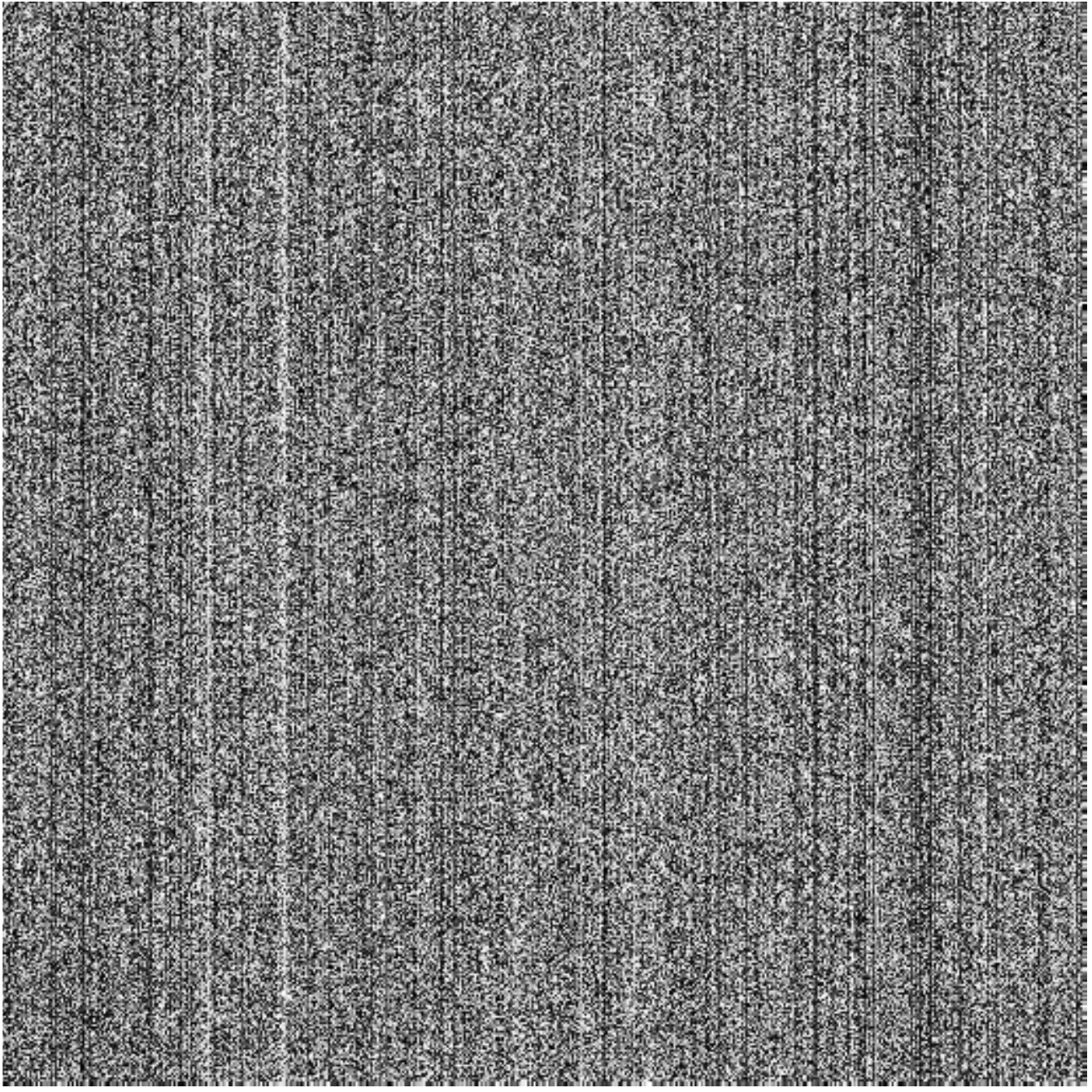}
b)
\end{minipage}
\caption{A pair of plain-image and the corresponding cipher-image:
a) image ``Lenna''; b) cipher-image of a).}
\label{fig:APairPlaintext}
\end{figure}

\begin{figure*}[!htb]
\centering
\begin{minipage}[t]{4.5\imagewidth}
\centering
\includegraphics[width=4.5\imagewidth]{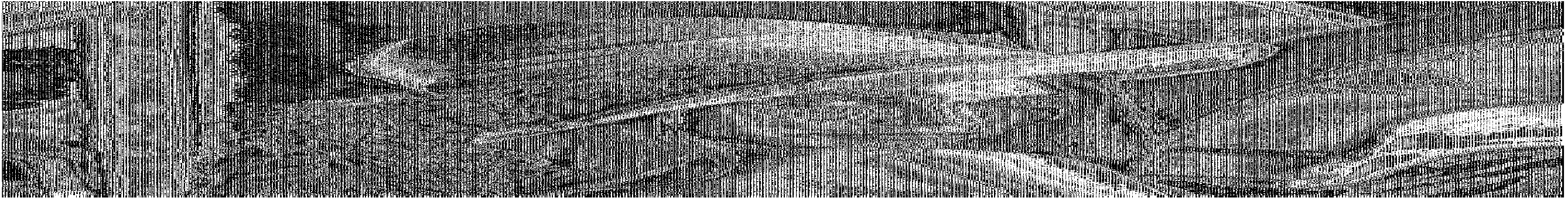}
a)
\end{minipage}
\begin{minipage}[t]{4.5\imagewidth}
\centering
\includegraphics[width=4.5\imagewidth]{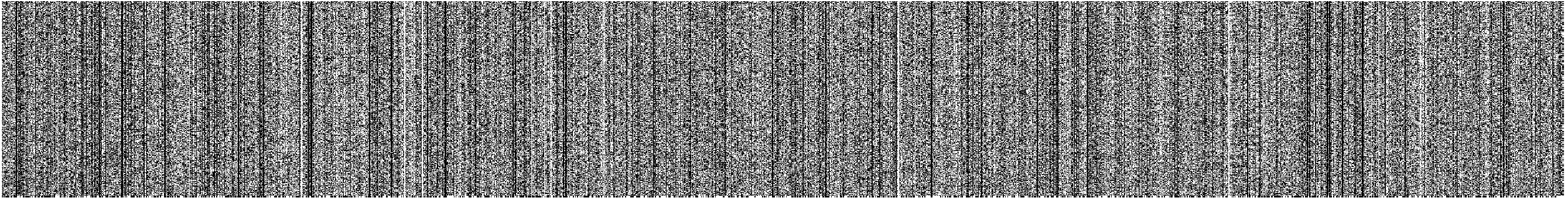}
b)
\end{minipage}
\begin{minipage}[t]{4.5\imagewidth}
\centering
\includegraphics[width=4.5\imagewidth]{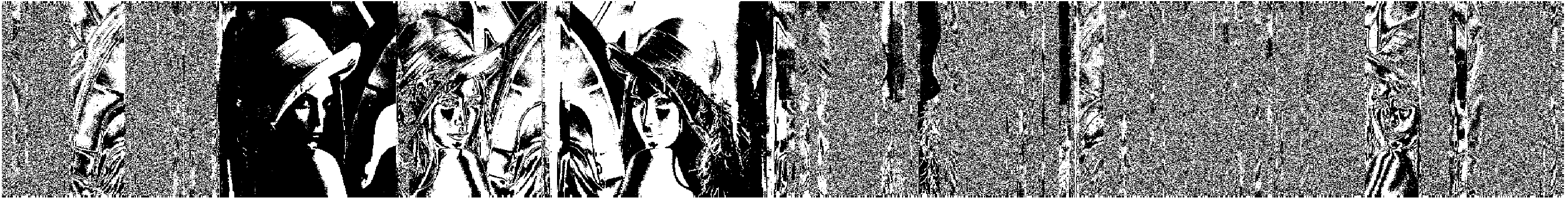}
c)
\end{minipage}
\caption{Some intermediate images for breaking ISEA: a) binary presentation of the image in Fig.~\ref{fig:APairPlaintext}a);
b) binary presentation of the image in Fig.~\ref{fig:APairPlaintext}b); c) the recovered binary image from b).}
\label{fig:BinImageCOA}
\end{figure*}

\begin{figure}[!htb]
\centering
\begin{minipage}[t]{\imagewidth}
\centering
\includegraphics[width=\imagewidth]{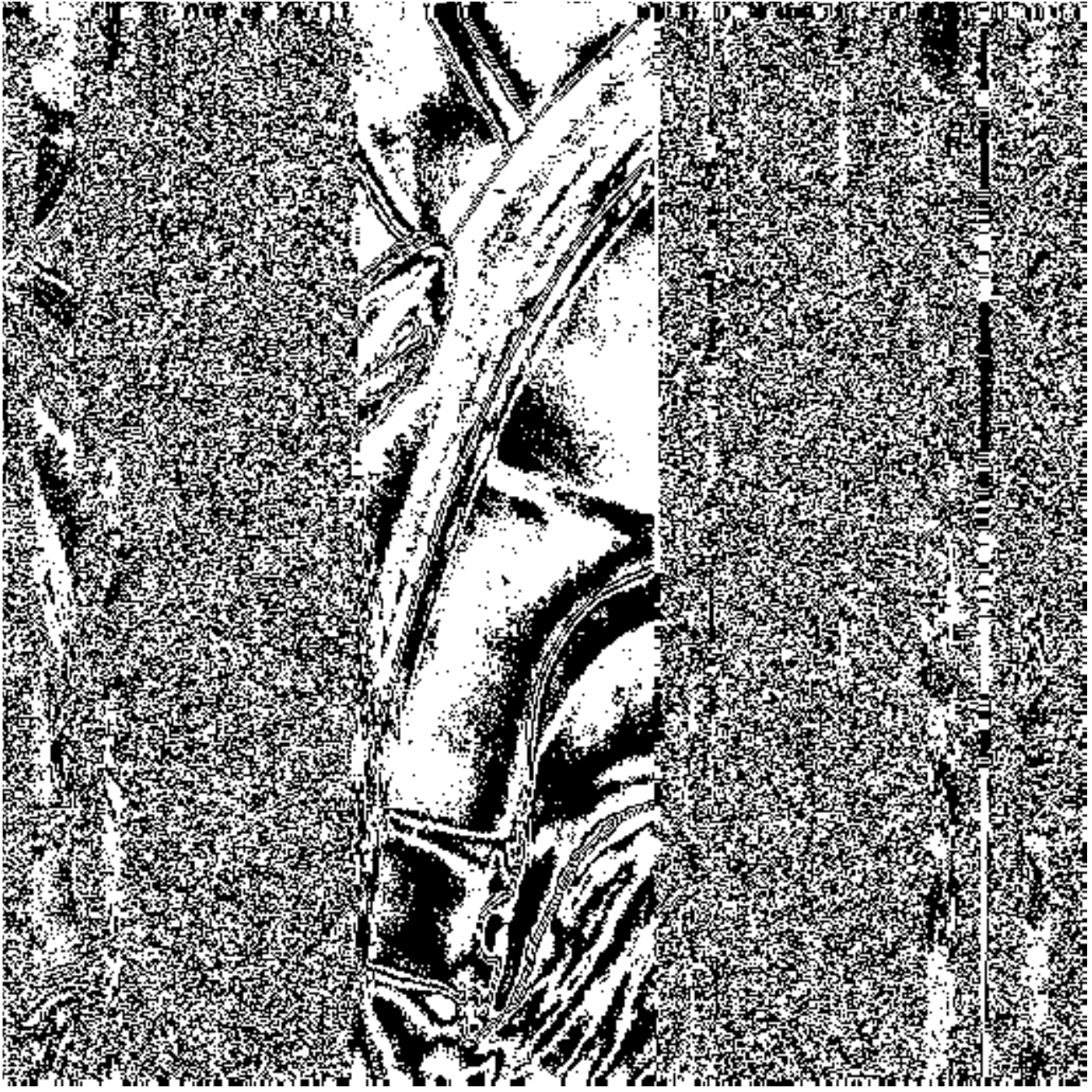}
a)
\end{minipage} \hspace{4pt}
\begin{minipage}[t]{\imagewidth}
\centering
\includegraphics[width=\imagewidth]{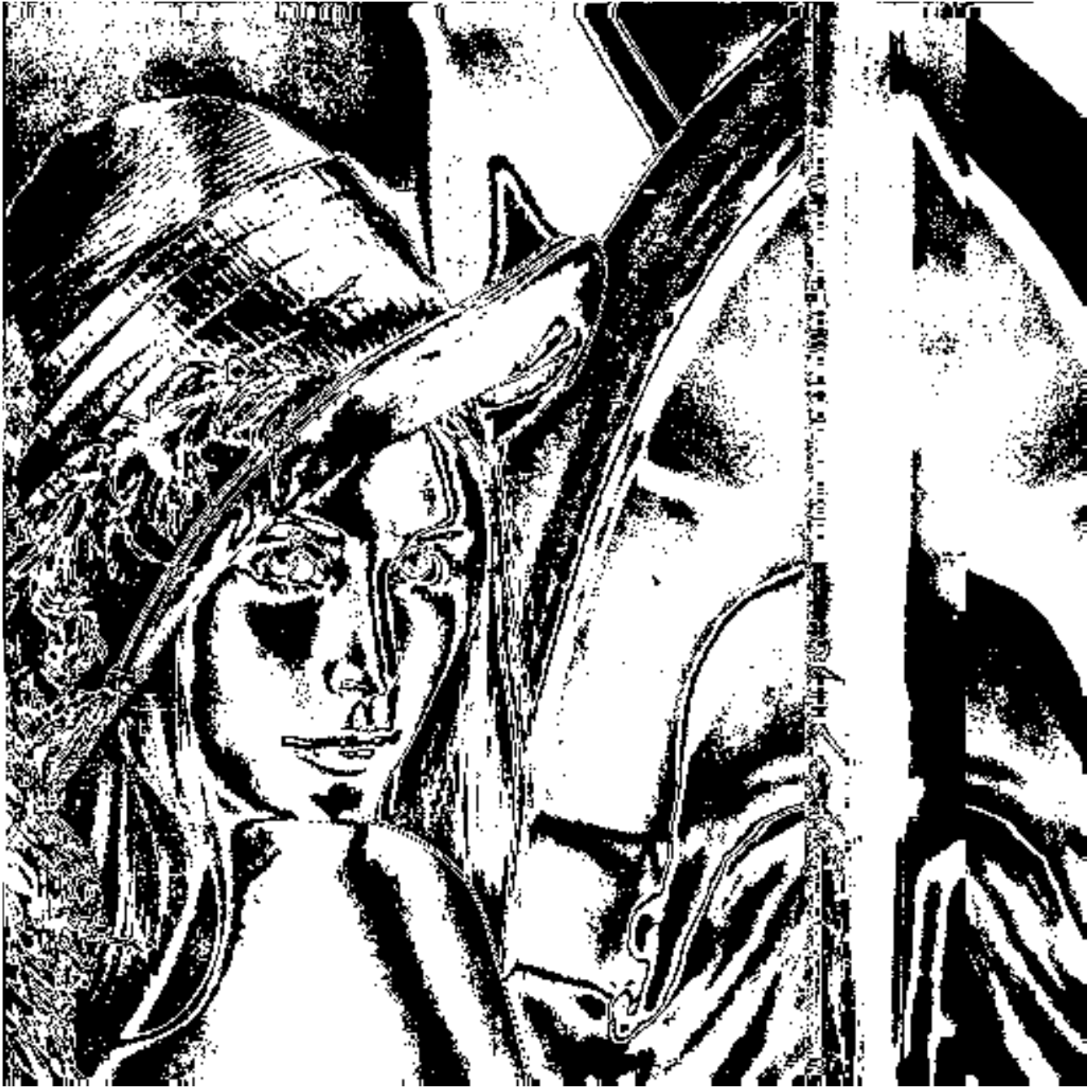}
b)
\end{minipage}
\\
\begin{minipage}[t]{\imagewidth}
\centering
\includegraphics[width=\imagewidth]{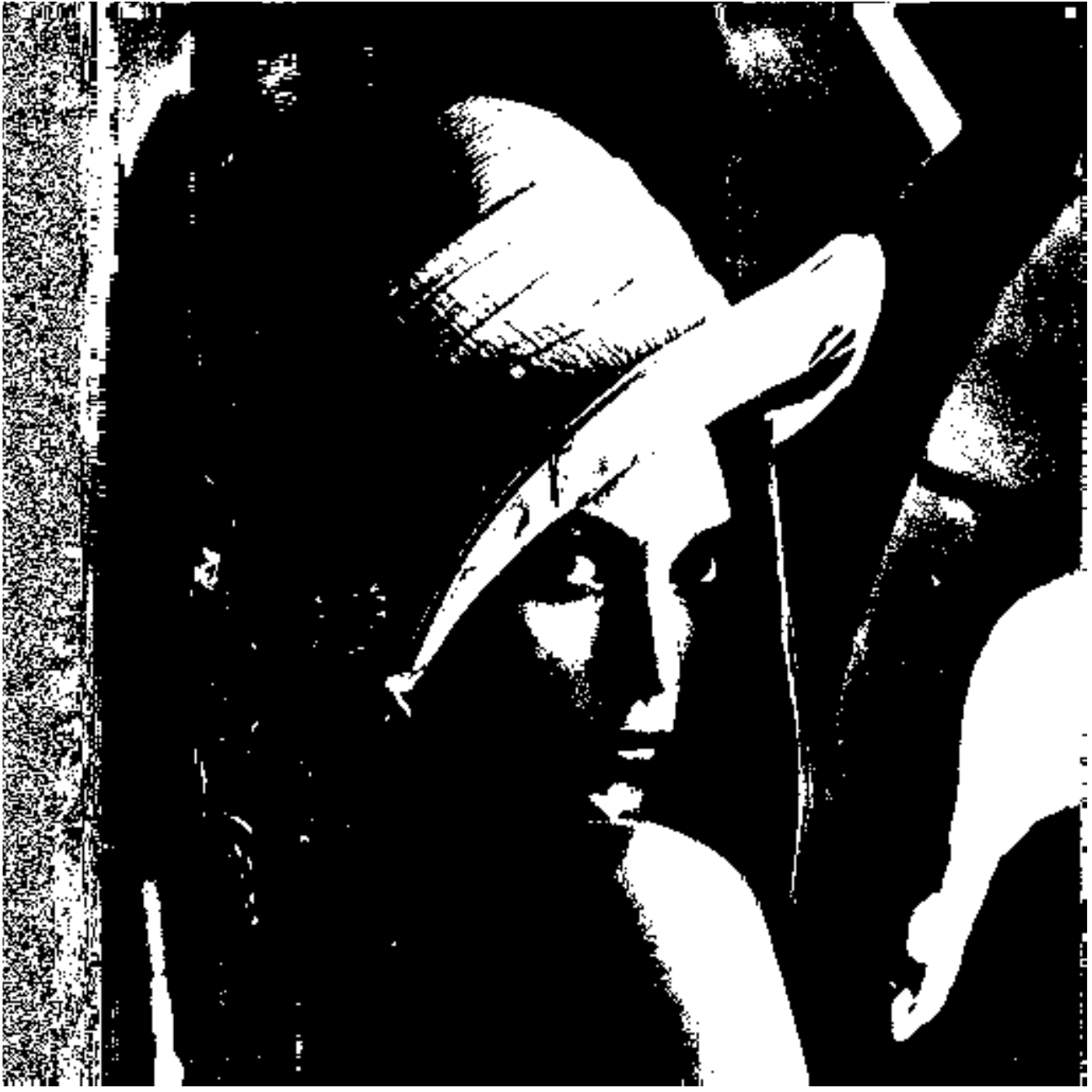}
c)
\end{minipage} \hspace{4pt}
\begin{minipage}[t]{\imagewidth}
\centering
\includegraphics[width=\imagewidth]{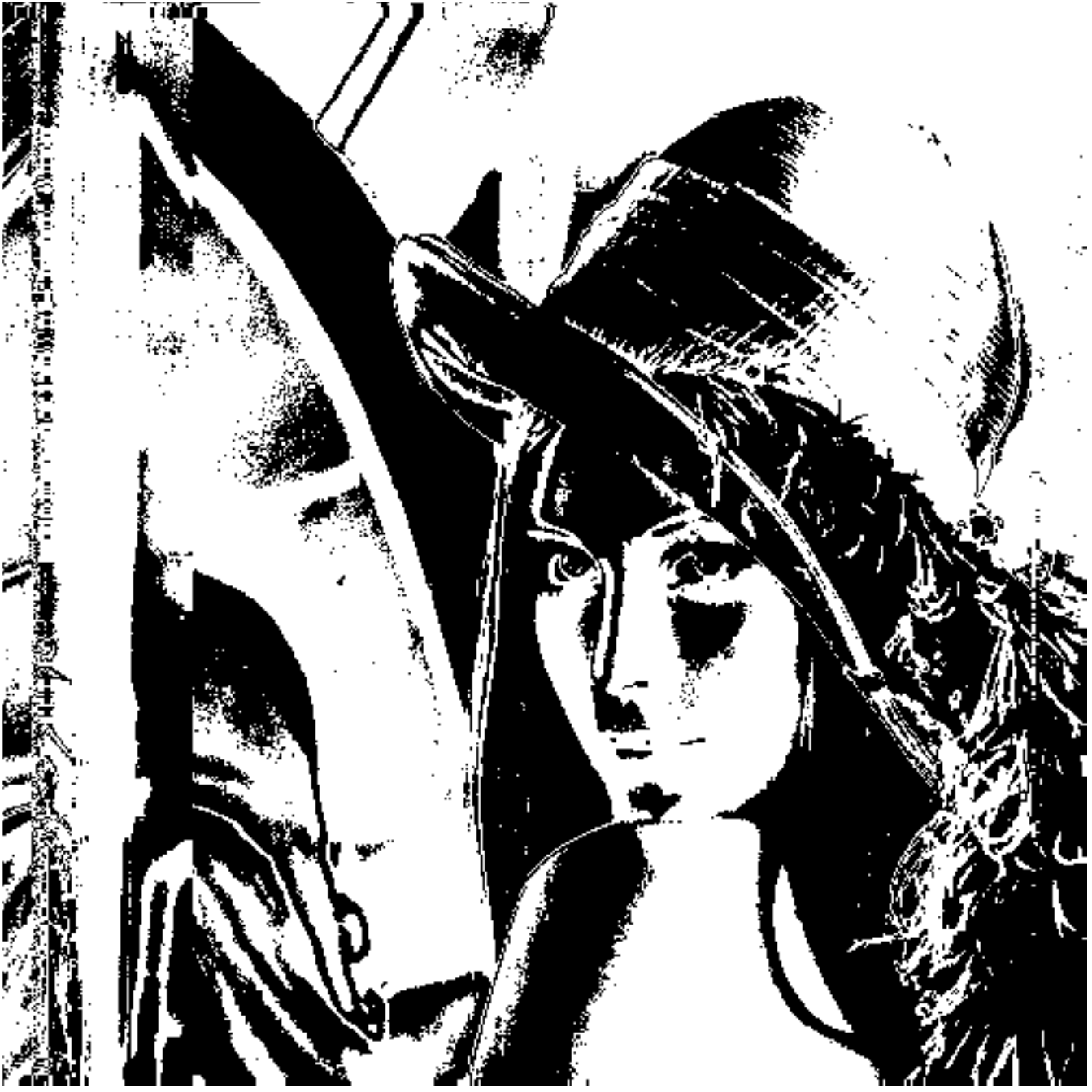}
d)
\end{minipage}
\caption{Four visible segments cropped from the image shown in Fig.~\ref{fig:BinImageCOA}c).}
\label{fig:coa2}
\end{figure}

Comparing Fig.~\ref{fig:BinImageCOA}a) and Fig.~\ref{fig:BinImageCOA}b), one can see that most of the visual information contained in Fig.~\ref{fig:BinImageCOA}a)
is concealed by the scrambling operations. However, the correlations existing between the rows and columns of image are not changed.
Given a row vector or a column vector of the cipher-image, its neighbouring vector may be found by searching for the vector possessing the highest
correlation index. Repeating this process iteratively in the horizontal and vertical directions, an approximate version of $\bm{B}$ can be obtained.
Actually, the vector search problem can be considered as one of the binary template matching problems discussed in \cite{TUBBS:binary:PR89}. For simplicity, we selected the ratio of the same bits existing in two binary vectors, namely Sokal and Michncr's measure discussed in \cite{TUBBS:binary:PR89}, as the similarity measure.
For example, the image shown in Fig.~\ref{fig:BinImageCOA}c) is the result obtained from that in Fig.~\ref{fig:BinImageCOA}b) by this approach. Some visible image blocks can be automatically detected with the image quality metric defined in \cite{Wang:ImageQuality:TIP04}, where the image composed of the eight bit-planes of the image in Fig.~\ref{fig:APairPlaintext}a) in the corresponding weight order is used as the reference. Four cropped
binary images are shown in Fig.~\ref{fig:coa2}, which reveals some important visual information, especially the rough sketch, of the original image shown in Fig.~\ref{fig:APairPlaintext}a). As shown in Fig.~\ref{fig:coa2}d), the vector order of the obtained result may be reversible to that of the right version since only the relative locations among the vectors can be confirmed.

\subsection{Known-plaintext attack on ISEA}

The known-plaintext attack is an attack model for cryptanalysis where the attacker know every implementation detail of the analyzed encryption algorithm (Shannon's axiom)
and can access to a set of plaintexts and the corresponding ciphertexts encrypted with the same unknown secret key. Because the two basic scrambling parts of ISEA are exerted in two
orthogonal directions, its ability to withstand the known-plaintext attack is weaker than other position permutation-only encryption algorithm costing encryption complexity of the same magnitude.

The equivalent version of the secret key of ISEA can be recovered by the following traditional method:
1) For each plain-bit, put the positions of the cipher-bits of the same value into the set containing its permutation position candidates;
2) If more than one sets of plain-images and the corresponding cipher-image are available, intersect the sets containing permutation-position candidates and
obtain a final set for each plain-bit;
3) Assign the estimated permutation position of each plain-bit with the first available (unassigned) element in its permutation position set.
Referring to \cite{Lcq:Optimal:SP11}, one can assure that $O(\lceil \log_2(8MN)\rceil)$ known plain-images can support efficient break of ISEA by considering it as a position-scrambling encryption algorithm exerting on
a scrambling domain of size $M\times 8N$ with permuted elements of two possible values. Essentially, ISEA is composed by cascading two basic permutation-only schemes of following parameters:
scrambling domain of size $M\times 1$ with permuted elements of $2^{(8N)}=256^N$ possible values; scrambling domain of size $1\times(8N)$ with permuted elements of $2^M$ possible values.
Due to the cascading, the two basic permutation-only schemes can not be broken separately in a direct way. However, some and even all parts of them can be recovered
iteratively. Concrete approaches are described as follows.
\begin{itemize}
\item \textit{Step 1}: Compare the number of elements 1's in each row vector of $\bm{B}'$ with that in all row vectors of $\bm{B}$.
If $\bm{B}(j^*, :)$ is the unique row of $\bm{B}$ possessing the same number of elements 1's as $\bm{B}'(i^*, :)$, one can assure
$\bm{T_M}(j^*)=i^*$ and put $i^*$ into set $\bm{R}$.

\item \textit{Step 2}: Similar to \textit{Step 1}, compare the number of elements 1's in each column vector of $\bm{B}'$ with that in all column vectors of $\bm{B}$.
If the former is unique in the latter, the corresponding element in $\bm{T_N}$ can be recovered correctly and put the column number into set $\bm{C}$
(the initial states of sets $\bm{R}$ and $\bm{C}$ are both empty).

\item \textit{Step 3}: Compare each column vector of $\{\bm{B}^*(i, :)\}_{i=i^*_1}^{i^*_r}$ with all column vectors of $\{\bm{B}'(i, :)\}_{i=i^*_1}^{i^*_r}$
and add its column number into set $\bm{C}$ if there exists the unique same column vector, where $\{\bm{B}^*(i, :)\}_{i=i^*_1}^{i^*_r}$ is recovered via Eq.~(\ref{eq:encrypt_ver}) and
$\bm{R}=\{i^*_1, \cdots, i^*_r\}$.

\item \textit{Step 4}: Similar to \textit{Step 3}, compare each row vector of $\{\bm{B}^*(:, j)\}_{j=j^*_1}^{j^*_c}$ with all row vectors of $\{\bm{B}(:, j)\}_{j=j^*_1}^{j^*_c}$
and add its row number into set $\bm{R}$ if there exists the unique same row vector, where $\{\bm{B}^*(:, j)\}_{j=j^*_1}^{j^*_c}$ is obtained via Eq.~(\ref{eq:encrypt_hor}),
and $\bm{C}=\{j^*_1, \cdots, j^*_c\}$.

\item \textit{Step 5}: Iteratively repeat \textit{Step 3} and \textit{Step 4} till the sizes of sets $\bm{R}$ and $\bm{C}$ are both not increased
(only the new-found number is added into $\bm{R}$ and $\bm{C}$ in the above steps).
\end{itemize}

If more pairs of known plain-images and the corresponding cipher-images encrypted with the same secret key are available,
the sets $\bm{R}$ and $\bm{C}$ can be further expanded with the above steps. The values of $\bm{T_M}(i^{**})$ and $\bm{T_N}(j^{**})$ can be assigned with the index of the first vector possessing the same measure (the number of elements 1's in a vector or the vector itself) during the above comparisons, where $i^{**}\in (\mathbb{Z}_{M}-\bm{R})$ and $j^{**}\in(\mathbb{Z}_{8N}-\bm{C})$.
As demonstrated in \cite{Li:Permutation:SPIC2008,Lcq:Optimal:SP11}, more known plain-images can make the vectors possessing the same
comparison measure less, and $\bm{T_M}(i^{**})$ and $\bm{T_M}(j^{**})$ can be correctly guessed in higher probability.
In addition, they can be further checked with the methods given in the previous subsection.

\begin{figure}[!htb]
\centering
\begin{minipage}[t]{0.7\imagewidth}
\centering
\includegraphics[width=0.7\imagewidth]{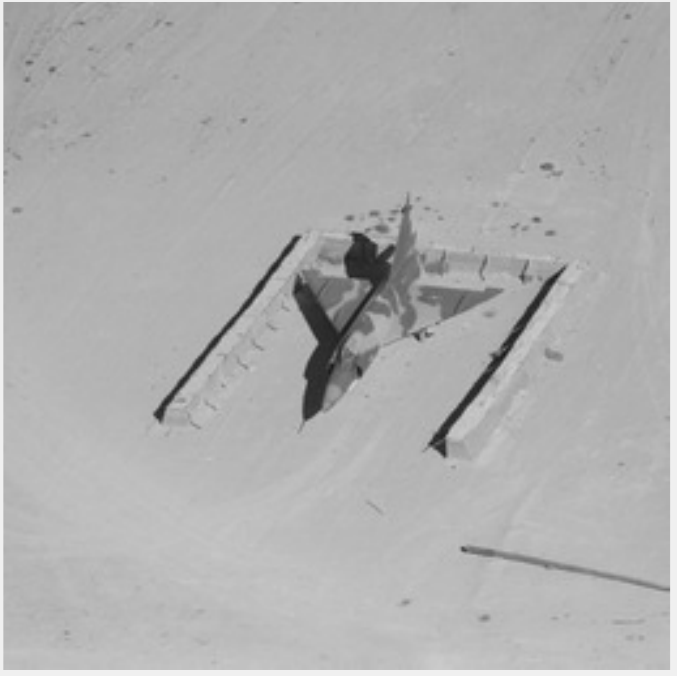}
a)
\end{minipage} \hspace{3pt}
\begin{minipage}[t]{0.7\imagewidth}
\centering
\includegraphics[width=0.7\imagewidth]{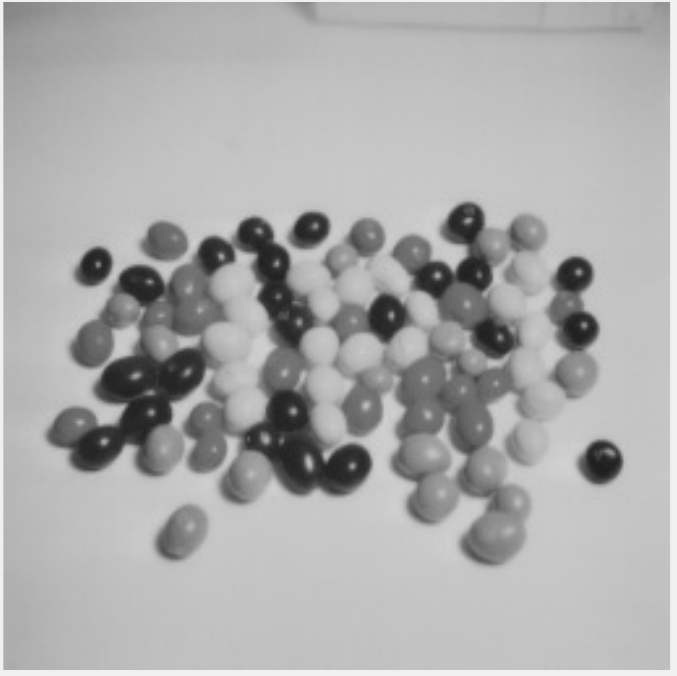}
b)
\end{minipage}
\hspace{3pt}
\begin{minipage}[t]{0.7\imagewidth}
\centering
\includegraphics[width=0.7\imagewidth]{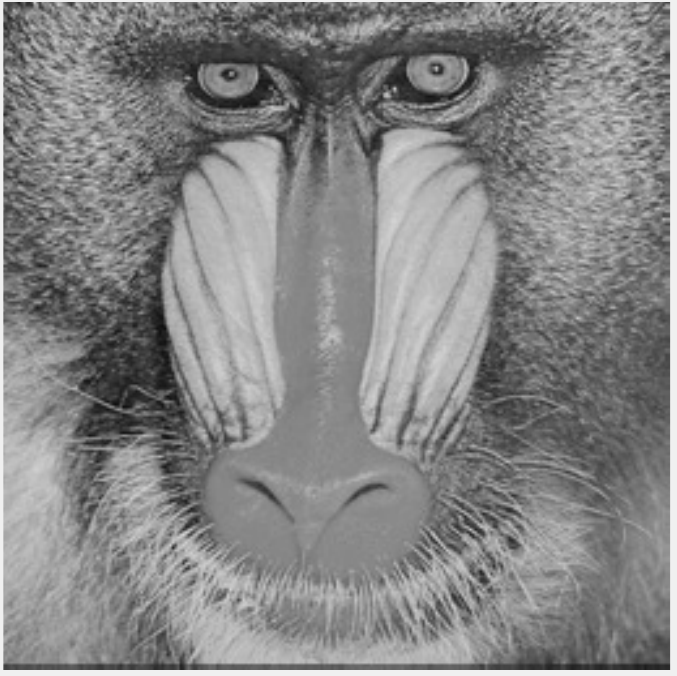}
c)
\end{minipage}\\
\begin{minipage}[t]{0.7\imagewidth}
\centering
\includegraphics[width=0.7\imagewidth]{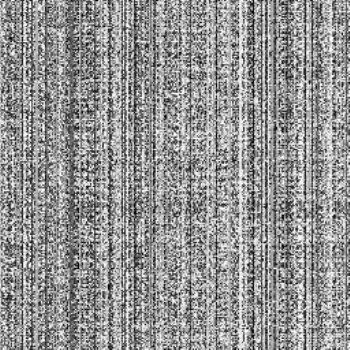}
d)
\end{minipage}
\hspace{3pt}
\begin{minipage}[t]{0.7\imagewidth}
\centering
\includegraphics[width=0.7\imagewidth]{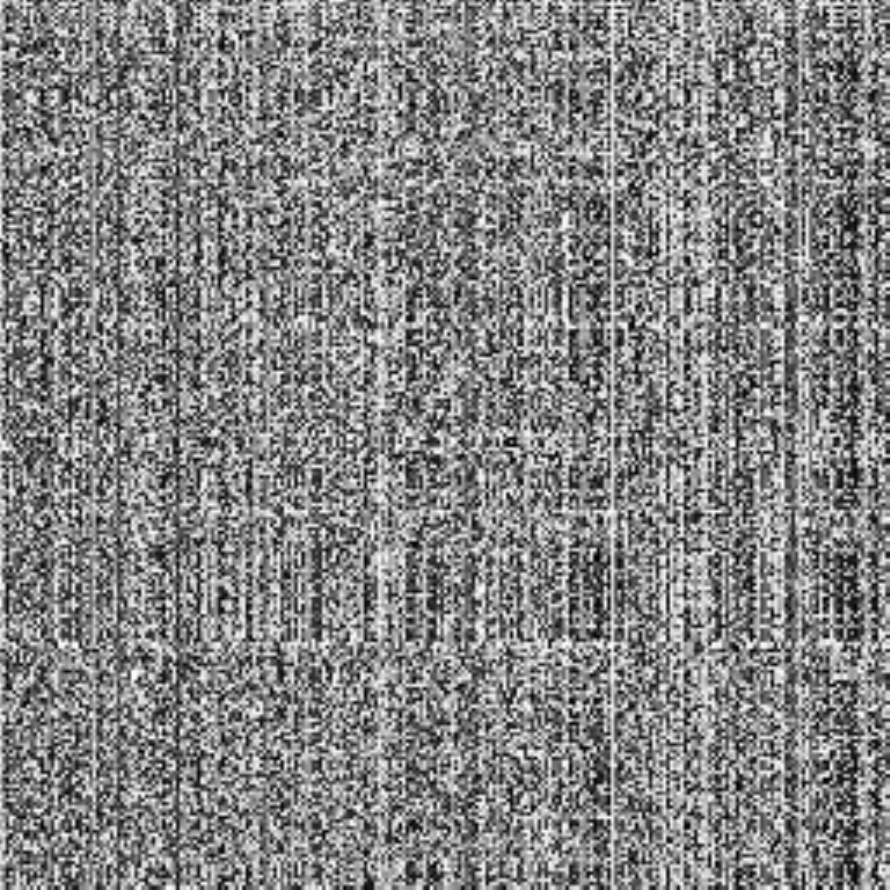}
e)
\end{minipage}
\hspace{3pt}
\begin{minipage}[t]{0.7\imagewidth}
\centering
\includegraphics[width=0.7\imagewidth]{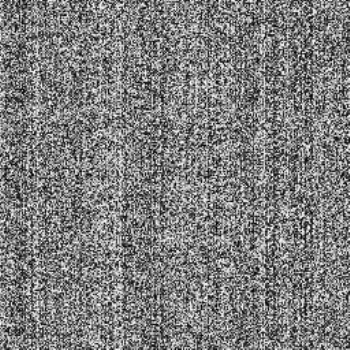}
f)
\end{minipage}
\caption{Three pairs of plain-image and the corresponding cipher-image: a) ``Fighter Jet''; b) ``Candy''; c) ``Baboon'';
d) encrypted ``Fighter Jet''; e) encrypted ``Candy''; f) encrypted ``Baboon.''}
\label{fig:3PairPlaintext}
\end{figure}

\begin{figure}[!htb]
\centering
\begin{minipage}[t]{2\imagewidth}
\centering
\includegraphics[width=2\imagewidth]{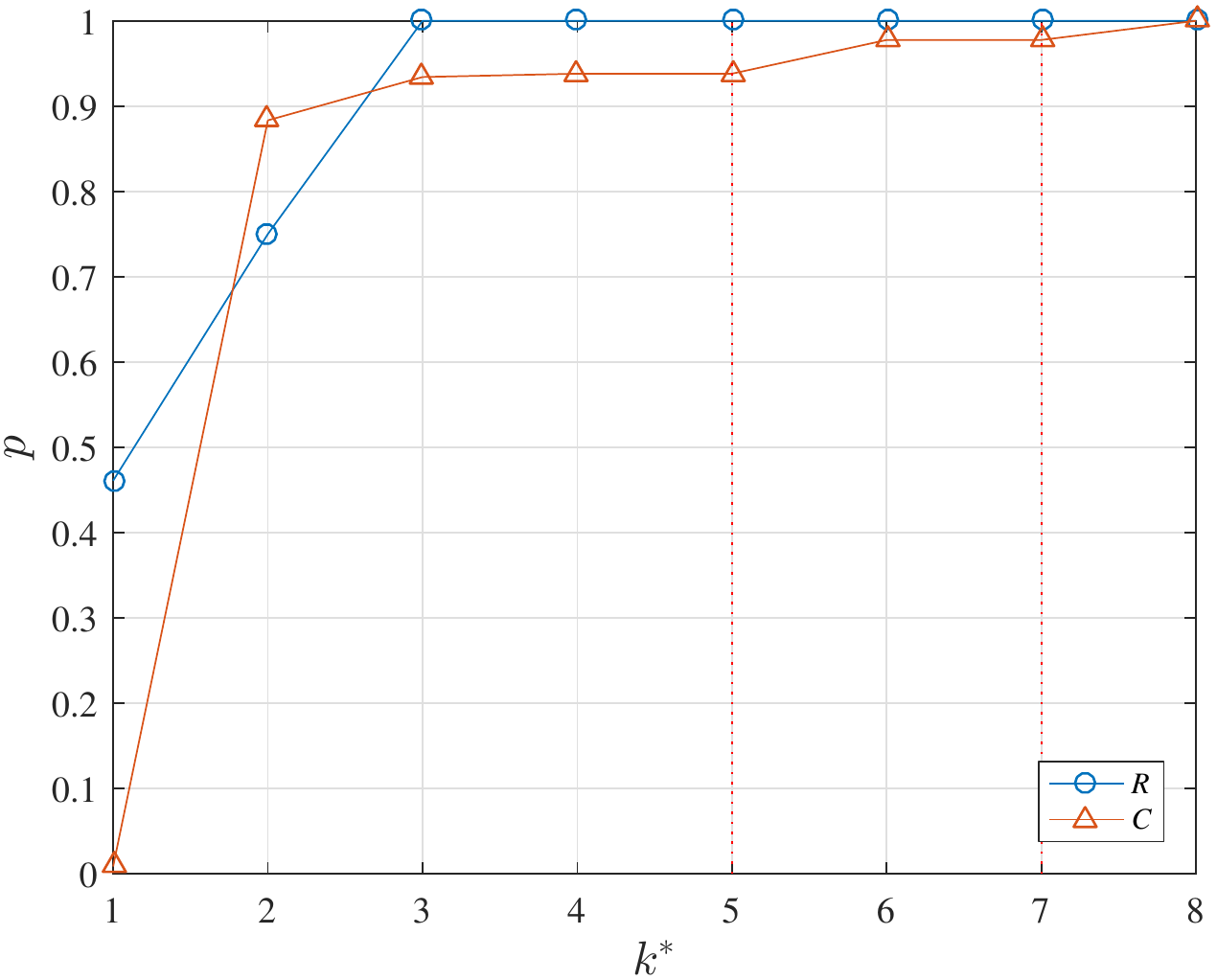}
\end{minipage}
\caption{The ratios between the sizes of $\bm{R}$, $\bm{C}$ and their maximal possible sizes during the attack with three pairs of plain-images,
where $k^*$ denotes the index of the attacking steps.}
\label{fig:ratio}
\end{figure}

To verify real performance of the attack given in this subsection, we illustrate it with three pairs of known plain-images of size $256\times 256$, shown in Fig.~\ref{fig:3PairPlaintext}, and the corresponding cipher-images encrypted with secret key $m=20$, $n=51$, $x_0=0.2009$, and $\mu=3.98$.
Using the known plain-image shown in Fig.~\ref{fig:3PairPlaintext}a), the sizes of $\bm{R}$ and $\bm{C}$ can reach 118 and 18 after \textit{Step 2}, respectively. Note that the size of $\bm{B}$ is $256\times 2048$. The size of $\bm{R}$ is further increased to 192 and 256 after the $1$-st and $2$-rd operation of \textit{Step 4}, respectively. In contrast, the size of $\bm{C}$ is only increased to 1809, 1913, 1921 after the 1-st, $2$-rd and $3$-th operation of \textit{Step 3}, respectively. As the $4$-th operation of \textit{Step 3} has no effect on increasing the size of $\bm{C}$, we used the second plain-image shown in Fig.~\ref{fig:3PairPlaintext}b), which increases it to 2002. Its maximal value was obtained with the third plain-image shown in Fig.~\ref{fig:3PairPlaintext}c).
The ratios between sizes of $\bm{R}$, $\bm{C}$ and their respective maximal values during the above attack processes are shown in
Fig.~\ref{fig:ratio}, which confirms the perfect performance of the proposed known-plaintext attack.

\subsection{Chosen-plaintext attack on ISEA}

The chosen-plaintext attack is an enhanced version of the known-plaintext attack, where an attacker can arbitrarily choose the plaintexts
to optimize the breaking performance, e.g. reduce the number of needed plaintexts, and/or increase the accuracy of the obtained information of the secret key.
In \cite{Zhao:scrambling:CNSNS2012}, a set of specific plain-images and the corresponding cipher-images were selected to accurately obtain the scrambling relation
in the two directions, respectively. However, we have found that the method can be further optimized, as some chosen plain-images can be used to break the scrambling operations in
the two orthogonal directions at the same time.

If $M\leq 8N$, the equivalent secret key of ISEA can be recovered with the following steps:
\begin{itemize}
\item \textit{Step 1}: Choose a plain-image satisfying $\bm{B}=[\bm{B}_L, \bm{B}_R]$, where
\newcommand\bigzero{\makebox(0,0){\text{\Large0}}}
\begin{equation}
\bm{B}_L=\bm{T}_L\cdot
\left[\begin{array}{c}
\begin{matrix}
1     &         &           &  \\
1     & 1       & \bigzero  &  \\
\vdots& \vdots  & \ddots    &  \\
1     & \cdots  & 1         & 1
\end{matrix}
\end{array}\right]_{M\times M}\cdot \bm{T}_R,
\label{eq:specialmatrix}
\end{equation}
$\bm{T}_L$ and $\bm{T}_R$ are any permutation matrices of size $M\times M$, and $\bm{B}_R$ is an $M\times (8N-M)$ matrix with constant elements 0's or 1's.

\item \textit{Step 2}: As the number of 1's in every row of $\bm{B}$ is different and the horizontal permutation has no influence on it,
one can exactly recover the vertical permutation vector $\bm{T_M}$ by comparing the number of 1's in each row of $\bm{B}$ and that of $\bm{B}'$.

\item \textit{Step 3}: Now, ISEA has degenerated to a permutation-only scheme operating in a scrambling domain of size $8N$ with permuted elements of $2^M$ possible values.
Referring to Eq.~(5) in \cite{Licq:hierarchical:SP2016}, one can ensure that the $8N$ elements of vector $\bm{T_N}$ can be exactly recovered with
$\lceil \log_{2^M}(8N)\rceil$ chosen plain-images, where
\[\bm{B}_k(i, j)=\lfloor j/2^{M\cdot k+i}\rfloor \bmod 2,\]
$\lfloor x \rfloor$ gives the largest integer less than or equal to $x$,
$k\in \{0, 1, \cdots, \lceil \log_{2^M}(8N)\rceil-1\}$, $i\in \{0, 1, \cdots, M-1\}$, and $j\in \{0, 1, \cdots, 8N-1\}$.
\end{itemize}

If $8N\in\{M, M+1\}$, the plain-image with $\bm{B}$ in (\ref{eq:specialmatrix}) can be used to accurately confirm $M$ elements of $\bm{T_N}$
with the same approach as in \textit{Step 2}. The sole remaining element of $\bm{T_N}$ can be recovered also when $8N=M+1$.
So, the number of required chosen plain-images for breaking ISEA is $\lceil \log_{2^M}(8N)\rceil=\lceil\frac{1}{M}(3+\log_{2}N)\rceil$ if $M<8N-1$;
and one if $8N\in\{M, M+1\}$. Similarly, one can deduce that the required number is $1+\lceil \log_{2^{8N}}(M)\rceil=\lceil\frac{1}{8N}(\log_{2}M)\rceil$ if $M>8N+1$; and one
if $M\in\{8N, 8N+1\}$. In all, one can conclude that the number of required chosen plain-images for breaking ISEA is
\begin{equation*}
n^*=1+
\begin{cases}
0                                    & \mbox{if } 8N\in\{M, M+1, M-1\};\\
\lceil\frac{1}{M}(3+\log_{2}N)\rceil & \mbox{if } 8N>M+1;\\
\lceil\frac{1}{8N}(\log_{2}M)\rceil  & \mbox{if } M>8N+1,
\end{cases}
\end{equation*}
which is much smaller than the estimated number
\begin{equation*}
n'=
\begin{cases}
\lceil 8N/M \rceil+1  & \mbox{if } M<N;\\
        9             & \mbox{if } M=N;\\
        \leq 9        & \mbox{if } 8N\ge M>N;\\
\lceil M/8N \rceil+1  & \mbox{if } M>8N,
\end{cases}
\end{equation*}
given in \cite[Sec. 5]{Zhao:scrambling:CNSNS2012}. Concretely, one has $n^*=2$ if $8N+1< M\leq 2^{8N}$ or $M+1< 8N\leq 2^M$.
As for an image of typical size $1704\times 2272$, one can calculate $n^*=1+\lceil\frac{1}{1704}(3+\log_{2}(2272))\rceil=2$ and $n'=\lceil 8\cdot 2272/1704 \rceil+1=11+1=12$.
As the computational load of the chosen-plaintext attack on a position permutation-only encryption algorithm is proportional to the product of the number of required plaintexts
by their sizes, the attacking complexity of the proposed method is $O(n^* \cdot MN)$ \cite{Li:Permutation:SPIC2008}.

\section{Conclusion}

In this paper, the security of a typical binary image scrambling encryption algorithm, called ISEA, against ciphertext-only attack and known/chosen-plaintext attacks
was studied comprehensively. Just as previous cryptanalytic works on the class of permutation-only encryption algorithms have shown, secret scrambling operations are incapable of providing a
sufficiently high level of security against known/chosen-plaintext attacks alone. Besides this, this cryptanalysis demonstrates the following added values on protecting multimedia data:
1) The correlation existing in multimedia data may be used to support some specific attacks and enhance breaking performance;
2) The size of each independent scrambling domain should be carefully checked to obtain the expected security requirement;
3) No matter what the permuted (scrambled) elements are, any permutation-only encryption algorithms will not change their histogram.
4) The security level of a multimedia encryption algorithm should be suitable for a given application scenario.

\section*{Acknowledgement}

This research was supported by the National Natural Science Foundation of China~(No.~61532020, 61611530549), the Hunan Provincial Natural Science Foundation of China~(No.~2015JJ1013), and the Scientific Research Fund of the Hunan Provincial Education Department~(No.~15A186).

\bibliographystyle{IEEEtran}
\bibliography{ISEA}

\vspace{10mm}



\noindent {\bf Chengqing Li} received M.Sc. degree in Mathematics from Zhejiang University, China in 2005 and Ph.D. degree in Electronic Engineering from City University of Hong Kong in 2008, respectively. He has been working at the College of Information Engineering, Xiangtan University, China as a Professor since 2016. His research interests include image privacy protection and multimedia cryptanalysis.
\\

\noindent {\bf Dongdong Lin} received his Bachelor degree in Computer Science from the College of Information Engineering, Xiangtan University in 2015, where he is currently pursing his Master degree in the same subject. His research interests include image privacy protection and image forensics.\\

\noindent {\bf Jinhu L\"u} received the Ph.D. degree in applied mathematics from the Academy of Mathematics and Systems Science (AMSS), Chinese Academy of Sciences, Beijing, in 2002. Currently, he is a Professor in the AMSS, Chinese Academy of Sciences. His research interests include chaotic dynamics, complex networks, and image forensics.
\end{document}